  \documentclass[twocolumn, aps, nofootinbib]{revtex4}           
\usepackage{amsmath}%
\usepackage{amsfonts}%
\usepackage{amssymb}%
\usepackage{graphicx}

\begin{document}
\title{Searching for a dilaton decaying to muon pairs at the LHC}
\author{Natascia Vignaroli}
\email{Natascia.Vignaroli@roma1.infn.it}
\affiliation{Dipartimento di Fisica, ``Sapienza" Universit\'{a} di Roma and \\ INFN Sezione di Roma, P.le A. Moro 2, I-00185 Roma, Italy}
\date{\today}


    \begin{abstract}
We analyze the decays to muons of a light dilaton produced via vector boson fusion at the LHC. We investigate models in which the electroweak symmetry breaking is triggered by a spontaneously broken, approximately conformal sector. Taking into account the possibility of shifts in the dilaton Yukawa couplings to muons, we find a rather promising scenario for the conformal model search in the channel, with the possibility for a dilaton discovery at a delivered luminosity of $100\ fb^{-1}$ at the LHC or, alternatively, for an extension of the exclusion zone in the model parameter space, until now fixed by the Tevatron. 
    \end{abstract}
 
\maketitle

\section{Introduction}
The decay to muons of a light dilaton at the LHC can be a privileged channel to search for new physics. 
In particular, we consider conformal models with dilaton \cite{dilatone1,dilatone2}, in which the electroweak symmetry breaking (EWSB) is triggered by a spontaneously broken, nearly conformal sector. The dilaton, $\chi$, is the pseudo-Goldstone boson associated to the spontaneous breaking of the conformal symmetry and has Higgs-like properties. When an explicit breaking of the conformal symmetry is introduced, dilaton couplings to light fermions, like muons, can receive significant enhancements that can notably improve the rate of a dilaton decaying to muons\footnote{The naturalness of this effect depends on the type of the explicit breaking. Significant enhancements for couplings to light fermions are more natural in the case of a breaking generated by operators involving fermions that are nearly marginal, as discussed in Ref. \cite{dilatone1}.}. The presence of enhancements in the dilaton couplings to light fermions is possible also when the dilaton is interpreted as the Randall-Sundrum Radion \cite{radion}; the analysis we perform is perfectly applicable to this case. \\
The dilaton Higgs-like properties allow us to start the analysis from the study of a light Higgs boson decaying to muons at the LHC. We retrace and optimize the analysis carried out in Ref. \cite{vbf_mumu}. This process profits by the distinguishing dimuon signature\footnote{The dimuon signal is very clean. Indeed, the LHC has a muon identification efficiency above $90\%$ \cite{cms_mu}.}, but, in the SM, it is disadvantaged by the low branching ratio (BR) for the Higgs decaying to muons. However, the EWSB mechanism in the SM and, in particular, the sector concerning the Higgs couplings to fermions is not explained and there are several hints, among which the excessive fine-tuning (hierarchy problem) for the Higgs mass, to suggest an EWSB mechanism different from the usual one based on a single Higgs doublet. The conformal model we investigate arises from these considerations. In the optics of possible variations from the SM in the Higgs couplings to fermions, the decay to muons of a light Higgs becomes a very interesting channel of analysis \footnote{In Refs. \cite{leptofilia, leptofilia2}, for example, an interesting scenario relative to a leptophilic Higgs in the 2HDM model is described. In this case, processes where the Higgs boson decays directly into a charged lepton pairs, like muon pairs, can contribute significantly to the discovery potential of a light Higgs at the LHC.} and studies for the decay to muons of a light Higgs produced via Vector Boson Fusion (VBF) \cite{vbf_mumu}, gluon fusion \cite{gg_mumu} and in association with $t\bar{t}$ pairs \cite{Su} have been carried out. The results of these studies show comparable significance values (about $2\sigma$, for an integrated luminosity of $300\ fb^{-1}$ at the LHC) for the three different production channels, in the Higgs mass range $115\ GeV\leq M_{H}\leq 120\ GeV$; on the contrary, for greater Higgs mass values, $120\ GeV< M_{H}\leq 140\ GeV$, the production in association with $t\bar{t}$ pairs display significance values smaller than those from the VBF and gluon fusion. Among the latter production channels, we choose to focus on the VBF. The VBF production, contrary to what happens for the gluon fusion, does not proceed via loop induced couplings to the dilaton. This allows a study less affected by the unknown contribution from heavy resonances in the loop induced couplings \cite{dilatone1,dilatone2}.\\
In this paper, we study, firstly, the decay to muons of a light Higgs boson produced, in the SM, via VBF at the LHC, checking and optimizing the analysis in Ref. \cite{vbf_mumu}. On the basis of the results obtained and taking into account theories beyond the SM predicting variations of the signal cross sections, we calculate the normalization factors $k=\frac{\left[\sigma(pp\rightarrow H)BR(H\rightarrow\mu^{+}\mu^{-})\right]_{NP}}{\left[\sigma(pp\rightarrow H)BR(H\rightarrow\mu^{+}\mu^{-})\right]_{SM}}$, needed to obtain a $3\sigma$ and a $5\sigma$ significance observation of the process, with an integrated luminosity of $100\ fb^{-1}$ at the LHC. Then, we focus on the conformal model with dilaton, taking into account the possibility of shifts in the dilaton Yukawa coupling to muons. The normalization factors are correlated to the parameters of this new physics model. So, we find the discovery regions in the parameter space, for an integrated luminosity of $100\ fb^{-1}$ at the LHC. We also take into account the Tevatron $95\%\ C.L.$ upper limit for the process $gg\rightarrow\chi\rightarrow\mu^{+}\mu^{-}$ cross section \cite{tev_coppie} and we derive exclusion zones for the conformal model parameters.

\section{Enhancement factors for the LHC}\label{k_factor}        
We study in this section the Higgs decaying to muons in VBF, $pp\rightarrow Hjj\rightarrow \mu^{+}\mu^{-}jj$, at the LHC with the center of mass energy $\sqrt{s}=14\ TeV$. The signal selection criteria we apply in this analysis are also valid for the dilaton. We find the enhancement factors to the signal cross section necessary to obtain a $3\sigma$ and a $5\sigma$ observation, with $100\ fb^{-1}$ at the LHC. We indicate such values as $k_{3\sigma}$ and $k_{5\sigma}$.
We start retracing the analysis carried out in Ref. \cite{vbf_mumu} with the aim to optimize some cuts and the valuation for renormalization and factorization scales. We focus on Higgs bosons with mass values in the range
\begin{equation}
	M_{H}\in\left[115\ GeV, 150\ GeV\right]\ .
	\label{higgs_mass}
\end{equation}
\noindent
The background we consider is the irreducible $\mu^{+}\mu^{-}jj$, where muons come from the decay $Z/\gamma^{*}\rightarrow\mu^{+}\mu^{-}$. This background has a QCD and an EW component. We ignore the reducible backgrounds: the $W^{+}W^{-}jj$ production, the $t\bar{t}+jets$ production and the $b\bar{b}jj$ production, that have negligible cross sections, once cuts are applied (like it is shown in Ref. \cite{vbf_mumu}). The cuts we use to reduce these backgrounds are the following:
\begin{gather}
|\eta_{j}|<4.5, \ |\eta_{j1}-\eta_{j2}|>4.2, \ \eta_{j1}\cdot\eta_{j2}<0, \ \Delta R_{jj}>0.6 \nonumber \\
 \  p_{Tj}\geq30\ GeV, \ m_{jj}\geq700\ GeV   \nonumber \\
p_{T\mu}\geq10\ GeV \ \textnormal{\footnotesize{and at least one muon with}} \ p_{T\mu}\geq20\ GeV  \nonumber \\
\  |\eta_{\mu}|<2.3, \ \Delta\eta_{\mu j}>0.6   \nonumber \\
\  M_{H}-1.6\ GeV<m_{\mu\mu}<M_{H}+1.6\ GeV\ .  
\label{set_b}
\end{gather}
\noindent
These cuts have been fixed taking into account the detector acceptances \cite{cms_mu} and exploiting the distinguishing features of the production via VBF \cite{vbf_rainw}. Indeed, the signal process is characterized by a particular topology, with two energetic jets, one forward and the other backward, emitted, on average, at high rapidity (the limit on the jet rapidity is fixed at the value $4.5$, taking into account the limitation on the hadronic calorimeter, $|\eta_{j}|<5$, and the spread of the jets cone) and with two isolated muons produced, on average, in the central rapidity region between the two tagging jets. Then, the invariant mass of the jets generated by VBF mechanism is, on average, higher than the invariant mass of the jets coming from QCD background. The cuts also include limits on the muons transverse momentum, that derive from trigger issues, and a limitation on the muons invariant mass, that we impose to be included in the mass region centered on the Higgs mass, $M_{H}\pm1.6\ GeV$, which is anticipated to capture $68\%$ of the signal cross section\footnote{The latter imposition is very effective in background rejection, because the backgrounds muons invariant mass has a peak on the Z mass, that is far away from the Higgs mass values we are taking into account.}(like estimated in \cite{vbf_mumu}). The cuts on eq. (\ref{set_b}) are the same cuts adopted in Ref. \cite{vbf_mumu} with the exception of the cuts on the jets transverse momentum, $p_{Tj}$, and on the jets invariant mass, $m_{jj}$, that we have better optimized. We have taken into account several $p_{Tj}$ limit values higher than the $20\ GeV$ value used in Ref. \cite{vbf_mumu}, in order to analyze the effects on signal-background ratios ($S/B$) and significances. Indeed, provided that significance and $S/B$ values do not suffer of remarkable reductions, it is preferable to have final jets with an high transverse momentum, because there is more distinction among signal and minimium bias events. We find that the limit on $p_{Tj}$ can be extended up to $30\ GeV$, without entailing reductions on significance values greater than $2\%$ and leading, instead, to increases up to $20\%$ on $S/B$ ratios. Moreover, we find that it is advantageous to extend the lower limit of $500\ GeV$ on $m_{jj}$, used in Ref. \cite{vbf_mumu}, up to $700\ GeV$; indeed, we find increases up to $35\%$ on $S/B$ ratios while the statistical significance remains the same.\\
In Tab. \ref{tab:finali} signal and backgrounds cross sections are reported for different values of Higgs boson mass. These values have been calculated by the event generator MadGraph/MadEvent \cite{madgraph}\footnote{We have taken into account the NLO width for the Higgs, obtained from the HDECAY program \cite{hdecay}.} at the factorization and renormalization scales
\begin{equation}
	\mu^{2}_{F}=\mu^{2}_{R}=p^{2}_{Tj1}+p^{2}_{Tj2}+p^{2}_{T\mu1}+p^{2}_{T\mu2}\ .
	\label{nuova_scala}
\end{equation}
\noindent
Adopting these scales, we can estimate, through the sum of muons square transverse momentum, the energetic contribution given to the process by the Higgs boson, for the signal, and the Z boson, for the backgrounds. In Tab. \ref{tab:finali} are also shown, for the different $M_{H}$ values, the signal-background ratios and the significance values, that are calculated for an integrated luminosity of $300\ fb^{-1}$. The significance is calculated as $\frac{S}{\sqrt{S+B}}$, where $S$ and $B$ respectively stand for signal and backgrounds number of events. In the calculation of significances and $S/B$ ratios we include the identification efficiencies of the final states. An efficiency value of $0.86$ is estimated for the identification of each jet and a value of $0.90$ for the identification of each muon \cite{vbf_mumu}. For the signal, a further $0.68$ factor is included, due to the constraint on the muons invariant mass. Besides the efficiencies, we also apply, as in Ref. \cite{vbf_mumu}, a minijet veto survival probability of $90\%$ for the signal ($\mathcal{V}_{S}$), $75\%$ for the EW background ($\mathcal{V}_{EW}$) and $30\%$ for the QCD background ($\mathcal{V}_{QCD}$). \\
Summarizing ($\epsilon_{S}$ and $\epsilon_{B}$ stand for signal and backgrounds efficiency), we have:

\begin{gather}
  \epsilon_{S}=40.8\% \ \ \epsilon_{B}=60\%  \nonumber \\
   \mathcal{V}_{S}=90\% \ \ \mathcal{V}_{EW}=75\% \ \ \mathcal{V}_{QCD}=30\% \ .
	\label{eff_veto}
	\end{gather}

\begin{figure}[ht]
\includegraphics{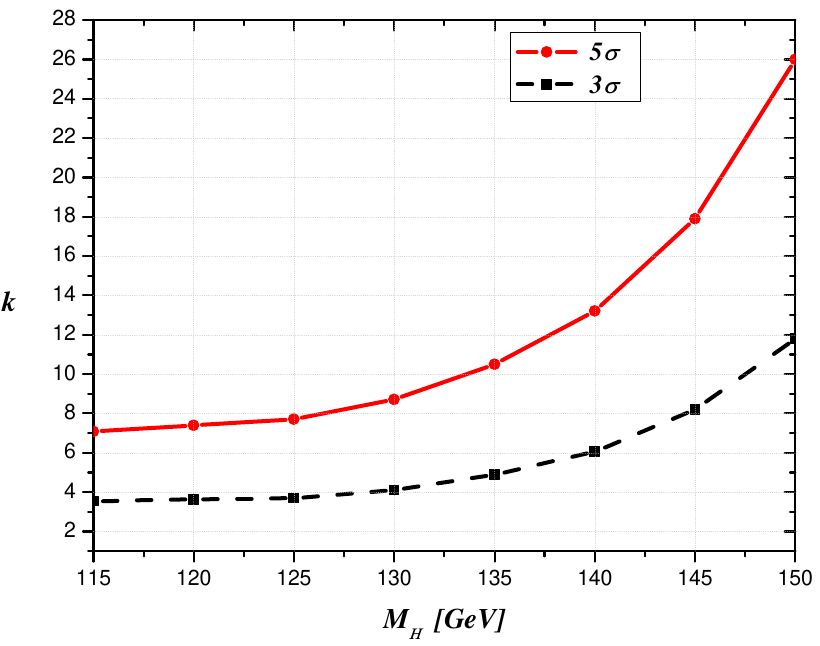}
	\caption{\small $k=\frac{\left[\sigma(pp\rightarrow H)BR(H\rightarrow\mu^{+}\mu^{-})\right]_{NP}}{\left[\sigma(pp\rightarrow H)BR(H\rightarrow\mu^{+}\mu^{-})\right]_{SM}}$ values necessary to obtain $3\sigma$ (dash line) and $5\sigma$ (continue line) significance, with an integrated luminosity of $100\ fb^{-1}$ at the LHC, for the process $pp\rightarrow Hjj\rightarrow\mu^{+}\mu^{-}jj$.}	
	\label{fig:fattori_k}	
\end{figure}

\begin{table}[ht]
	\centering
		\begin{tabular}{cccccc}
		\hline
	\multicolumn{6}{|c|}{$p_{Tj}>30\ GeV,\ m_{jj}>700\ GeV$}\\
	\multicolumn{6}{|c|}{$\mu^{2}_{F}=\mu^{2}_{R}=p^{2}_{Tj1}+p^{2}_{Tj2}+p^{2}_{T\mu1}+p^{2}_{T\mu2}$}\\
	\multicolumn{6}{|c|}{}\\
	  \hline
		\hline 
		\multicolumn{1}{|c|}{$M_{H}$}&
		\multicolumn{1}{@{}c@{}|}{$\sigma_{H}$}&
		\multicolumn{1}{@{}c@{}|}{$\sigma^{QCD}_{Z}$}&
		\multicolumn{1}{@{}c@{}|}{$\sigma^{EW}_{Z}$}&
		\multicolumn{1}{@{}c@{}|}{  $S/B$  }&
		\multicolumn{1}{@{}c@{}|}{$\frac{S}{\sqrt{S+B}}$}\\
		\multicolumn{1}{|c|}{$[GeV]$}&
		\multicolumn{1}{@{}c@{}|}{$[fb]$}&
	  \multicolumn{1}{@{}c@{}|}{$[fb]$}&
	  \multicolumn{1}{@{}c@{}|}{$[fb]$}&
	   \multicolumn{1}{@{}c@{}|}{}&	
	  \multicolumn{1}{@{}c@{}|}{}\\
		\hline
		\multicolumn{1}{|c|}{115} & \multicolumn{1}{|c|}{0.157} & \multicolumn{1}{|c|}{0.879} & \multicolumn{1}{|c|}{0.221} & \multicolumn{1}{|c|}{0.224} & \multicolumn{1}{|c|}{1.78}\\
		\hline
		\multicolumn{1}{|c|}{120} & \multicolumn{1}{|c|}{0.143} & \multicolumn{1}{|c|}{0.760} & \multicolumn{1}{|c|}{0.170} & \multicolumn{1}{|c|}{0.246} & \multicolumn{1}{|c|}{1.76}\\
		\hline
	  \multicolumn{1}{|c|}{125} & \multicolumn{1}{|c|}{0.127} & \multicolumn{1}{|c|}{0.527} & \multicolumn{1}{|c|}{0.140} & \multicolumn{1}{|c|}{0.296} & \multicolumn{1}{|c|}{1.79}\\
		\hline 
		\multicolumn{1}{|c|}{130} & \multicolumn{1}{|c|}{0.108} & \multicolumn{1}{|c|}{0.441} & \multicolumn{1}{|c|}{0.115} & \multicolumn{1}{|c|}{0.302} & \multicolumn{1}{|c|}{1.66}\\
		\hline
		\multicolumn{1}{|c|}{135} & \multicolumn{1}{|c|}{0.0877} & \multicolumn{1}{|c|}{0.402} & \multicolumn{1}{|c|}{0.101} & \multicolumn{1}{|c|}{0.273} & \multicolumn{1}{|c|}{1.44}\\
		\hline
		\multicolumn{1}{|c|}{140} & \multicolumn{1}{|c|}{0.0679} & \multicolumn{1}{|c|}{0.322} & \multicolumn{1}{|c|}{0.0990} & \multicolumn{1}{|c|}{0.243} & \multicolumn{1}{|c|}{1.21}\\
		\hline
		\multicolumn{1}{|c|}{145} & \multicolumn{1}{|c|}{0.0493} & \multicolumn{1}{|c|}{0.294} & \multicolumn{1}{|c|}{0.0955} & \multicolumn{1}{|c|}{0.189} & \multicolumn{1}{|c|}{0.928}\\
		\hline
		\multicolumn{1}{|c|}{150} & \multicolumn{1}{|c|}{0.0334} & \multicolumn{1}{|c|}{0.283} & \multicolumn{1}{|c|}{0.0804} & \multicolumn{1}{|c|}{0.141} & \multicolumn{1}{|c|}{0.673}\\
		\hline
		\end{tabular}
\caption{\small Results of the $pp\rightarrow Hjj\rightarrow \mu^{+}\mu^{-}jj$ analysis. Signal and background cross sections are calculated at the factorization and renormalization scales of eq. (\ref{nuova_scala}) and with the application of the set of cuts in eq. (\ref{set_b}). Significance $S/\sqrt{S+B}$ and signal-background ratio values refer to an integrated luminosity of $300\ fb^{-1}$. To such values the efficiency and the minijet veto factors reported in eq. (\ref{eff_veto}) are applied.}
\label{tab:finali}   
\end{table}
\noindent
Our results confirm substantially the results in Ref. \cite{vbf_mumu}. We exploit them to give predictions for theory beyond the SM and then, in particular, for the conformal model with dilaton. \\
In Fig. \ref{fig:fattori_k} the $k_{3\sigma}$ and $k_{5\sigma}$ values, obtained on the basis of the results for $pp\rightarrow Hjj\rightarrow \mu^{+}\mu^{-}jj$ analysis in the SM (Tab. \ref{tab:finali}), are shown as functions of $M_{H}$.
We can observe a growing trend for the enhancement factors in the $M_{H}$ range, $115\ GeV \leq M_{H}\leq 150\ GeV $. Signal cross section increases of about one order of magnitude are sufficient to obtain a $5\sigma$ significance, with an integrated luminosity of $100\ fb^{-1}$ at the LHC. More in detail, for $M_{H}$ values in the range $115\ GeV \leq M_{H}\leq 140\ GeV $ an enhancement factor of $k\sim 3.5-6$ is needed for a $3\sigma$ signal observation and $k\sim 7-13$ for a $5\sigma$ signal observation. Then, for $M_{H}$ values in the range $140\ GeV \leq M_{H}\leq 150\ GeV $, higher $k_{3\sigma}$ and $k_{5\sigma}$ values are required: $k\sim 6-12$ for a $3\sigma$ signal observation and $k\sim 13-26$ for a $5\sigma$ signal observation.    

\section{Conformal model with dilaton}
In this section we apply the $k_{5\sigma}$ factors to the new physics model described in Refs. \cite{dilatone1,dilatone2}, that predicts the presence of a light CP even scalar, the dilaton, with Higgs-like properties. This model rises from the consideration that the EWSB mechanism can be different from the single Higgs doublet mechanism in the SM. The EWSB can be, for example, triggered by a spontaneously broken nearly conformal sector. The spontaneous breaking of the conformal symmetry occurs at an energy scale $f\geq v$, where $v$ is the electroweak scale, $v\cong246\ GeV$. The pseudo-Goldstone boson connected with the spontaneous breaking is the dilaton $\chi$, which has couplings to the SM fields of the same form of the Higgs boson ones\footnote{Indeed, the Higgs boson can be seen as a dilaton, in the limit case $f=v$.}. Then, a small explicit breaking of the conformal symmetry is introduced\footnote{The symmetry breaking perturbation consists of a term $\lambda\mathcal{O}$, where $\lambda$ is a coupling and $\mathcal{O}$ is an operator of scaling dimension $\Delta\neq 4$. The explicit breaking is small either for $\lambda\ll f$ or for $\mathcal{O}$ nearly marginal, $\left|\Delta-4\right|\ll 1$ (Ref. \cite{dilatone1}).}. As a consequence, the dilaton acquires a mass and its couplings to the SM fermions can undergo variations from the previous form; in particular, when scale symmetry is violated by operators involving fermions, shifts in the dilaton Yukawa couplings to fermions can appear. 
Because of the smallness of the explicit breaking, the dilaton mass is naturally light and can be written as
\begin{equation}
	M^{2}_{\chi}=\epsilon f^{2}\ ,  
	\label{eq.cft.1}
\end{equation}
where $\epsilon$ is the parameter that controls deviations from exact scale invariance. The light dilaton couplings to the SM fields are parametrized in terms of the following low energy effective Lagrangian: 

\begin{gather}
\mathcal{L}_{\chi}=-\frac{\bar{\chi}}{f}\sum_{\psi}\left(m_{\psi}+\epsilon y^{(1)}_{\psi}v\right)\bar{\psi}\psi \nonumber \\
+\left(2\frac{\bar{\chi}}{f}+\frac{\bar{\chi}^{2}}{f^{2}}\right)\left(M^{2}_{W}W^{+}_{\mu}W^{-\mu}+\frac{1}{2}M^{2}_{Z}Z_{\mu}Z^{\mu}\right) \nonumber \\
+\frac{\bar{\chi}}{f}\left[\frac{\alpha}{8\pi}c_{EM}\left(F_{\mu\nu}\right)^{2}+\frac{\alpha_{s}}{8\pi}c_{G}\left(G_{\mu\nu}\right)^{2}\right]\ ,
		\label{eq.lagrangiana}
\end{gather}
\noindent
$\bar{\chi}$ in eq. (\ref{eq.lagrangiana}) is the dilaton physical field, canonically normalized, that describes the fluctuations from the $vev$ $\left\langle \chi\right\rangle=f$ and $y^{(1)}$ are $3\times 3$ diagonal matrices in flavor space. The Lagrangian in eq. (\ref{eq.lagrangiana}) has the same form of the Higgs one in the SM, the main differences are in the presence of a scale $f$, instead of the electroweak scale $v$, and in the presence of shifts in the dilaton Yukawa couplings to fermions, $y^{(1)}_{\psi}$. Then, the relevant parameters for our phenomenological study are: the dilaton mass, $M_{\chi}$, the ratio $v/f$ between the electroweak scale, $v\cong246\ GeV$, and the scale of the conformal spontaneous breaking, $f\geq v$, the shifts in the dilaton Yukawa couplings to fermions, $y^{(1)}_{\psi}$, the dilaton loop induced couplings to massless gauge bosons, $c_{EM}$, for the photons, and $c_{G}$, for the gluons. These couplings depend sensitively on assumptions about ultraviolet physics\footnote{The choice to focus on the VBF production, instead of gluon fusion, will allow to obtain predictions in the parameters space, $(y^{(1)}_{\mu}, v/f)$, less sensitive to $c_{G}$ variations.}. \\ 
Because of the presence of the shifts $y^{(1)}_{\psi}$, the ratio between the partial width for the dilaton decaying to fermions and the analogous width for the Higgs in the SM depends on a factor
\begin{equation}
 \left(\frac{v}{f}\right)^{2}\left(\frac{m_{\psi}+\epsilon y^{(1)}_{\psi}v}{m_{\psi}}\right)^{2} \ .
	\label{eq.cft.30}	
\end{equation}
So, depending on $y^{(1)}_{\psi}$ values\footnote{Large shifts are more natural in the case of a breaking generated by the addition of a nearly marginal operator to the conformal Lagrangian; otherwise fine tunings with fermionic masses are required, as discussed in Ref. \cite{dilatone1}.}, there can be significant increases in the partial widths for the decays to light fermions. Instead, because of the presence of the scale $f\geq v$, cross sections for the dilaton production, in all channels, are suppressed by a factor $(v/f)^{2}$ with respect to the cross sections for the Higgs production. Therefore, signal cross sections are affected by a suppression to the VBF, given by a factor $(v/f)^{2}$. However, this suppression can be compensated by the possible increase of the BR for the light dilaton decaying to muons. In the following we take into account two different possibilities, corresponding to the presence of shifts only in the dilaton Yukawa coupling to muons and of analogous shifts for all the dilaton Yukawa couplings to leptons.

\section{Discovery regions for the dilaton}
\begin{figure}[ht]
\scalebox{0.94}{\includegraphics{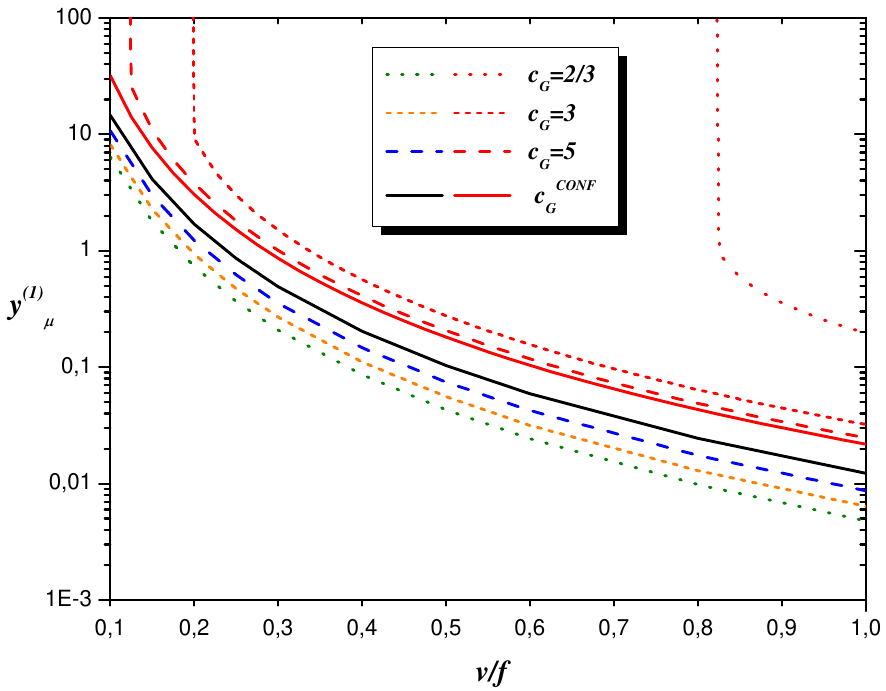}}	
		\caption{\small Discovery regions for the process $pp\rightarrow\chi jj\rightarrow\mu^{+}\mu^{-}jj$ with a dilaton of $M_{\chi}=150\ GeV$, for an integrated luminosity of $100\ fb^{-1}$ at the LHC. We assume the presence of a shift only for the dilaton Yukawa coupling to muons, $y^{(1)}_{\mu}\neq 0$ and $y^{(1)}_{\psi\neq\mu}=0$. Different regions are shown, in correspondence to different gluons-dilaton couplings, $c_{G}$ ($c^{CONF}_{G}=23/3$). The lower curves refer to the $y^{(1)}_{\mu}$ values necessary to obtain a $5\sigma$ significance, with an integrated luminosity of $100\ fb^{-1}$ at the LHC; the upper curves denote, instead, the exclusion zones, which are obtained from the Tevatron. The possible discovery regions extend, for each $c_{G}$ value, from the lower curves to the upper curves.}	
	\label{y_vf_cg}
\end{figure}

We now take into account the possible presence of shifts in the dilaton Yukawa couplings to muons, $y^{(1)}_{\mu}\neq 0$ ($y^{(1)}_{\psi\neq\mu}=0$), and we find, for different gluons-dilaton couplings $c_{G}$, the values of the parameters $v/f$ and $y^{(1)}_{\mu}$ necessary to obtain a $5\sigma$ significance for the process $pp\rightarrow\chi jj\rightarrow\mu^{+}\mu^{-}jj$, with an integrated luminosity of $100\ fb^{-1}$ at the LHC. \\
We consider a $c_{G}$ parameter variation that goes from a minimum value of $2/3$, which corresponds to the case of a SM-like coupling, to a value $c^{CONF}_{G}=23/3$, which corresponds to the case of a QCD completely embedded in the conformal sector. We neglect, instead, the possible $c_{EM}$ variation from its SM value, which has a negligible influence on the BR for the dilaton decaying to muons\footnote{BR for the dilaton decaying to photons cannot exceed the percent level.}. We make use of the $k_{5\sigma}$ values, found in \ref{k_factor}. We can relate these values to the conformal model parameters, considering that the variation of the signal cross section from the Higgs signal in the SM can be expressed by the ratio\footnote{We assume the same backgrounds for the dilaton and the Higgs signals.}:
\begin{equation}
	k=\frac{\sigma_{\chi}BR_{\chi}(\mu)}{\sigma_{H}BR_{H}(\mu)}=\left(\frac{v}{f}\right)^{2}\frac{BR_{\chi}(\mu)}{BR_{H}(\mu)}\ ,
	\label{eq.ratio}
\end{equation}
\noindent
where $\sigma_{\chi}$($\sigma_{H}\equiv\sigma_{SM}$) is the cross section for the dilaton (Higgs) production and $BR_{\chi}(\mu)$ ($BR_{H}(\mu)$) is the BR for the dilaton (Higgs) decaying to muons. The ratio $BR_{\chi}(\mu)/BR_{H}(\mu)$ depends on the conformal model parameters; for example, in the case of a shift only for the muons and $c_{G}=2/3$, the $k$ ratio becomes
\begin{equation}
\left(\frac{v}{f}\right)^{2}\frac{\left(1+\frac{y^{(1)}_{\mu}}{m_{\mu}}\frac{M^{2}_{\chi}}{v}\left(\frac{v}{f}\right)^{2}\right)^{2}}{1+BR_{H}(\mu)\left(1+\frac{y^{(1)}_{\mu}}{m_{\mu}}\frac{M^{2}_{\chi}}{v}\left(\frac{v}{f}\right)^{2}\right)^{2}} = k(M_{\chi})\ .
	\label{eq.k_parametri}
\end{equation}
\noindent
We indicate as $y^{5\sigma}_{\mu}$ the values of the $y^{(1)}_{\mu}$ parameters that give a $5\sigma$ significance, with an integrated luminosity of $100\ fb^{-1}$ at the LHC.\\
In Tab. \ref{y_m}, we show the $y^{5\sigma}_{\mu}$ values, for different $c_{G}$ and $v/f$ values, in function of the dilaton mass. We find a quite soft dependence of the $y^{5\sigma}_{\mu}$ values on dilaton mass values, within the range $[115\ GeV,150\ GeV]$; instead, there is a strong dependence on $v/f$ parameter. We show in Fig. \ref{y_vf_cg}, referred to a dilaton with $M_{\chi}=150\ GeV$, the $y^{5\sigma}_{\mu}$ values, for different dilaton-gluons couplings $c_{G}$, as a function of the $v/f$ parameter. For the different $c_{G}$ values, the regions above the lower curves of the plot represent the discovery regions (significance values higher than $5\sigma$) for the process $pp\rightarrow\chi jj\rightarrow\mu^{+}\mu^{-}jj$, with an integrated luminosity of $100\ fb^{-1}$ at the LHC. Fig. \ref{y_vf_cg} shows that greater $y^{(1)}_{\mu}$ shifts are needed for the $5\sigma$ observation, when smaller $v/f$ and greater $c_{G}$ values are taken into account. This trend reflects the fact that signal cross sections are disadvantaged by $v/f$ reductions, because the dilaton production is suppressed by a factor $(v/f)^{2}$ compared to the Higgs production, and by $c_{G}$ increases, because $BR_{\chi}(\mu)$ decreases when the width for the dilaton decaying to gluons increases. The upper curves of the plot denote the exclusion regions for the model parameters. These regions can be obtained considering the Tevatron lepton pairs data \cite{tev_coppie}. Indeed, the not discovery at the Tevatron for the process $gg\rightarrow\chi\rightarrow\mu^{+}\mu^{-}$ fix an upper limit on the process cross section from which we derive upper limits on the conformal model parameters. On the basis of the results shown in Fig. \ref{y_vf_cg}, it is predictable that, in the case of a QCD completely embedded in the conformal sector ($c^{CONF}_{G}$), the LHC can extend the exclusion zone down to the underside curve related to a $5\sigma$ significance, or it can discover a signal in correspondence of the possible discovery region. Then, in the cases of smaller $c_{G}$ values, the possible discovery region expands and it becomes quite wide when the dilaton coupling to gluons is like the SM one ($c_{G}=2/3$). 

\begin{table}
	\centering
	\scalebox{0.75}{	
	\begin{tabular}{ccccccccc}\\
		\hline
		\multicolumn{1}{|c|}{} & \multicolumn{2}{|c|}{$y^{5\sigma}_{\mu}$}& \multicolumn{2}{|c|}{$y^{5\sigma}_{\mu}$}& \multicolumn{2}{|c|}{$y^{5\sigma}_{\mu}$}& \multicolumn{2}{|c|}{$y^{5\sigma}_{\mu}$}\\
\multicolumn{1}{|c|}{} & \multicolumn{2}{|c|}{}& \multicolumn{2}{|c|}{}& \multicolumn{2}{|c|}{}& \multicolumn{2}{|c|}{}\\		
		\multicolumn{1}{@{}|c|@{}}{$M_{\chi}$} & \multicolumn{2}{|c|}{$c_{G}=2/3$}& \multicolumn{2}{|c|}{$c_{G}=3$}& \multicolumn{2}{|c|}{$c_{G}=5$}& \multicolumn{2}{|c|}{$c_{G}=23/3$}\\
				\cline{2-9}
\multicolumn{1}{@{}|c|@{}}{$[GeV]$} & \multicolumn{1}{|c|}{$v/f$}& \multicolumn{1}{|c|}{$v/f$}& \multicolumn{1}{|c|}{$v/f$}& \multicolumn{1}{|c|}{$v/f$}& \multicolumn{1}{|c|}{$v/f$}& \multicolumn{1}{|c|}{$v/f$ }& \multicolumn{1}{|c|}{$v/f$}& \multicolumn{1}{|c|}{$v/f$}\\
		\multicolumn{1}{|c|}{} & \multicolumn{1}{|c|}{\ $0.1$\ }& \multicolumn{1}{|c|}{\ $0.3$\ }& \multicolumn{1}{|c|}{\ $0.1$\ }& \multicolumn{1}{|c|}{\ $0.3$\ }& \multicolumn{1}{|c|}{\ $0.1$\ }& \multicolumn{1}{|c|}{\ $0.3$\ }& \multicolumn{1}{|c|}{\ \ $0.1\ \ \ $}& \multicolumn{1}{|c|}{$0.3$}\\		
		\hline
		\multicolumn{1}{@{}|c|@{}}{115} & \multicolumn{1}{|c|}{5.71}& \multicolumn{1}{|c|}{0.179}& \multicolumn{1}{|c|}{9.00}& \multicolumn{1}{|c|}{0.290}& \multicolumn{1}{|c|}{13.1}& \multicolumn{1}{|c|}{0.430}& \multicolumn{1}{|c|}{19.2}& \multicolumn{1}{|c|}{0.634}\\
		\hline
	\multicolumn{1}{@{}|c|@{}}{120} & \multicolumn{1}{|c|}{5.35}& \multicolumn{1}{|c|}{0.168}& \multicolumn{1}{|c|}{8.46}& \multicolumn{1}{|c|}{0.273}& \multicolumn{1}{|c|}{12.3}& \multicolumn{1}{|c|}{0.405}& \multicolumn{1}{|c|}{18.0}& \multicolumn{1}{|c|}{0.598}\\	
		\hline
		\multicolumn{1}{@{}|c|@{}}{130} & \multicolumn{1}{|c|}{4.92}& \multicolumn{1}{|c|}{0.156}& \multicolumn{1}{|c|}{7.58}& \multicolumn{1}{|c|}{0.247}& \multicolumn{1}{|c|}{11.0}& \multicolumn{1}{|c|}{0.363}& \multicolumn{1}{|c|}{15.9}& \multicolumn{1}{|c|}{0.532}\\
		\hline
   \multicolumn{1}{@{}|c|@{}}{140} & \multicolumn{1}{|c|}{5.22}& \multicolumn{1}{|c|}{0.168}& \multicolumn{1}{|c|}{7.46}& \multicolumn{1}{|c|}{0.245}& \multicolumn{1}{|c|}{10.4}& \multicolumn{1}{|c|}{0.347}& \multicolumn{1}{|c|}{14.9}& \multicolumn{1}{|c|}{0.500}\\ 
    \hline
  	\multicolumn{1}{@{}|c|@{}}{150} & \multicolumn{1}{|c|}{6.40}& \multicolumn{1}{|c|}{0.209}& \multicolumn{1}{|c|}{8.14}& \multicolumn{1}{|c|}{0.269}& \multicolumn{1}{|c|}{10.7}& \multicolumn{1}{|c|}{0.355}& \multicolumn{1}{|c|}{14.6}& \multicolumn{1}{|c|}{0.491}\\
    \hline		
		\end{tabular}}
	\caption{\small $y^{(1)}_{\mu}$ values necessary to obtain a $5\sigma$ significance, with an integrated luminosity of $100\ fb^{-1}$ at the LHC ($y^{5\sigma}_{\mu}$), as functions of $M_{\chi}$, for different $c_{G}$ and $v/f$ values (we assume $y_{\psi\neq\mu}=0$). }
		\label{y_m}
\end{table}

\begin{figure}[ht]
	\centering
\scalebox{0.94}{\includegraphics{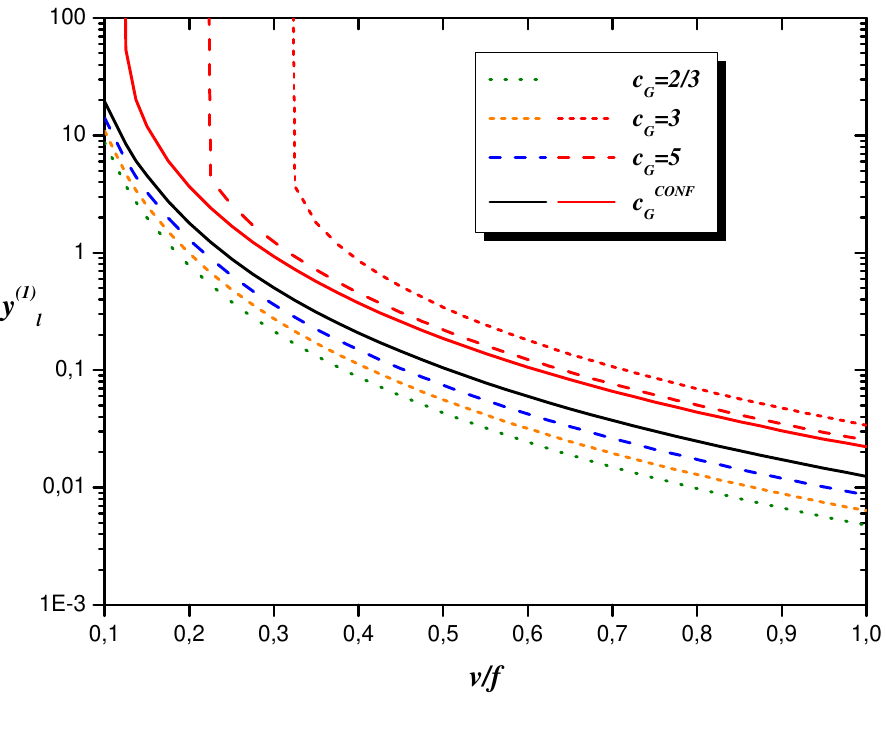}}
		\caption{\small Possible discovery regions for the process $pp\rightarrow\chi jj\rightarrow\mu^{+}\mu^{-}jj$ with a dilaton of $M_{\chi}=150\ GeV$ when the presence of analogous shifts for all the dilaton Yukawa couplings to leptons is assumed, $y^{(1)}_{\mu}=y^{(1)}_{\tau}=y^{(1)}_{e}$ and $y^{(1)}_{\psi\neq l}=0$.}		
		\label{y_lim_emutau}
\end{figure}

Finally, we have considered the possibility of analogous shifts for all the dilaton Yukawa couplings to leptons, that is $y^{(1)}_{\mu}=y^{(1)}_{\tau}=y^{(1)}_{e}$ and $y^{(1)}_{\psi\neq l}=0$\footnote{This is a valid assumption if, for example, a leptophilic scenario is hypothesized.}. Even when we consider, in this less advantageous assumption, the most unfavorable case that corresponds to a maximum $c_{G}$ value, $c^{CONF}_{G}=23/3$, we can obtain at the LHC, like Fig. \ref{y_lim_emutau} shows, significance values greater than $5\sigma$ or an extension, down to the underside curve related to a $5\sigma$ significance, of the exclusion zone. In the cases of smaller $c_{G}$ values, the possible discovery region expands (Tevatron lepton pairs data do not fix an exclusion zone for $c_{G}=2/3$). 

\section{Conclusions}
In this paper, we have shown that the decay to muons of a light CP even scalar produced via VBF can be an interesting channel of study at the LHC. Considering new physics models, we have shown that, in the Higgs mass range, $115\ GeV\leq M_{H}\leq 150\ GeV$, signal cross section increases of about one order of magnitude are sufficient to obtain a $5\sigma$ significance, with an integrated luminosity of $100\ fb^{-1}$ at the LHC. Then, a specific study for the conformal model with dilaton has been carried out. We have taken into account two different possibilities, corresponding to the presence of shifts only in the dilaton Yukawa coupling to muons and of analogous shifts for all the dilaton Yukawa couplings to leptons. In both cases, we have obtained a rather promising scenario for the conformal model search in the channel $\chi\rightarrow\mu^{+}\mu^{-}$ via VBF at the LHC, with the possibility for a dilaton discovery at a delivered luminosity of $100\ fb^{-1}$ or, alternatively, for an extension of the exclusion zone in the model parameter space, until now fixed by the Tevatron. Our study could be extended considering the other relevant production channels, gluon fusion and the production associated with $t\bar{t}$ pairs. Comparative studies for the conformal model in these different production channels could allow an estimate for model parameters, like the dilaton-gluons effective coupling $c_{G}$.   

\begin{acknowledgments}
I am especially grateful to B. Mele and S. Petrarca for having assisted me during the thesis and in this work. I thank F. Maltoni for suggestions and for hospitality at the Center for Particle Physics and Phenomenology (CP3) of the Catholic University of Louvain (Belgium), where part of this work was realized. I also thank R. Frederix, M. Herquet, C. Dhur and all the CP3 team for crucial assistance on Madgraph and R. Contino for comments on the manuscript.
\end{acknowledgments}





\end{document}